\documentclass{article}

     \usepackage[final,nonatbib]{neurips_mlsb_2020}

\usepackage[utf8]{inputenc} 
\usepackage[T1]{fontenc}    
\usepackage{hyperref}       
\usepackage{url}            
\usepackage{booktabs}       
\usepackage{amsfonts}       
\usepackage{nicefrac}       
\usepackage{microtype}      
\usepackage{xcolor}
\usepackage{graphicx}
\usepackage{amsmath}

\def\R{{\mathbb{R}}}

\title{Exploring generative atomic models in \\cryo-EM reconstruction}

\author{
  Ellen D. Zhong \\
  MIT \\
  \texttt{zhonge@mit.edu} \\
  \And
  Adam Lerer \\
  Facebook AI Research \\
  \texttt{alerer@fb.com} \\
  \And
  Joseph H. Davis \\
  MIT \\
  \texttt{jhdavis@mit.edu} \\
    \And
  Bonnie Berger \\
  MIT \\
  \texttt{bab@mit.edu} \\
}

\begin{document}

\maketitle

\begin{abstract}
    Cryo-EM reconstruction algorithms seek to determine a molecule's 3D density map from a series of noisy, unlabeled 2D projection images captured with an electron microscope.
    Although reconstruction algorithms typically model the 3D volume as a generic function parameterized as a voxel array or neural network, the underlying atomic structure of the protein of interest places well-defined physical constraints on the reconstructed structure. 
    In this work, we exploit prior information provided by an atomic model to reconstruct distributions of 3D structures from a cryo-EM dataset. 
    We propose Cryofold, a generative model for a continuous distribution of 3D volumes based on a coarse-grained model of the protein's atomic structure, with radial basis functions used to model atom locations and their physics-based constraints.  
    Although the reconstruction objective is highly non-convex when formulated in terms of atomic coordinates (similar to the protein folding problem), we show that gradient descent-based methods can reconstruct a continuous distribution of atomic structures when initialized from a structure within the underlying distribution.
    This approach is a promising direction for integrating biophysical simulation, learned neural models, and experimental data for 3D protein structure determination.

\end{abstract}

\section{Introduction}

Single particle cryo-electron microscopy (cryo-EM) is a powerful experimental technique for structure determination of proteins and macromolecular complexes at near-atomic resolution \cite{Nogales:2016bv}. In this technique, an electron microscope is used to image a purified sample of the molecule of interest suspended in vitreous ice. Initial processing of the resulting micrograph produces a dataset of $10^{4-7}$ 2D projection images, where each image captures a unique molecule suspended in a random, unknown orientation. Reconstruction algorithms are then used to computationally infer the 3D protein structure from the dataset. Unlike structure determination through x-ray crystallography, cryo-EM is able to determine the structure of molecules that are not amenable to crystallization, including those that adopt an ensemble of  heterogeneous conformational states. Thus, cryo-EM is in principle capable of modeling the entire ensemble of conformational states in a sample and probing the dynamic conformational landscape of proteins. To date, this promise has been limited by a lack of computational techniques for dealing with heterogeneous cryo-EM data, but promising new tools have recently been developed to reconstruct heterogeneous structures \cite{cryodrgn_nmeth, 3dva, multibody}.

\begin{figure}[t]
\centering
\includegraphics[width=300px]{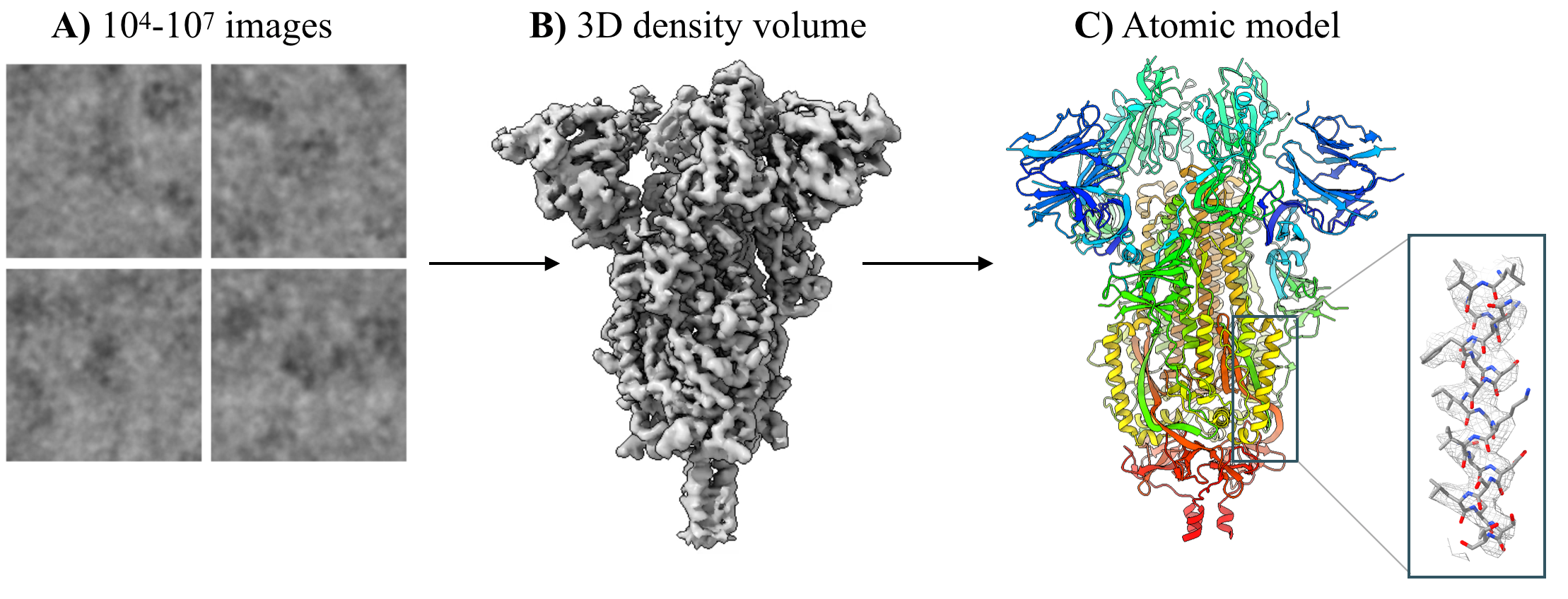}
\caption{\small{Structure determination via cryo-EM. Schematic of cryo-EM reconstruction (\textbf{A} $\rightarrow$ \textbf{B}) and atomic model fitting (\textbf{B} $\rightarrow$ \textbf{C}). In this work, we investigate the possibility of fitting the atomic protein structure directly during the reconstruction process. Dataset from Walls et al. \cite{walls2020structure}}.}
\label{fig:overview}
\end{figure}

Computational processing of cryo-EM datasets consists of several distinct stages (Figure \ref{fig:overview}). First, the noisy micrograph image is preprocessed and segmented, where bounding boxes containing individual molecules (i.e. particles) are identified and extracted (i.e. particle picking) \cite{topaz}.
Next, a 3D cryo-EM density volume (or distribution of volumes) that maps the molecule's electron scattering potential is reconstructed from the 2D projections \cite{cryodrgn_nmeth,cryosparc,relion3,cistem,eman2}. 
Finally, an atomic model is built into the resulting 3D volume either \textit{de novo} or fit starting from a related model, typically with the help of automated tools \cite{coot,phenix,mdff,Kihara:2020cf,Si:2020eg,isolde}. The reader is referred to \cite{Singer2020} for further details on the single particle cryo-EM image processing pipeline. 

In this work, we investigate the possibility of fitting the atomic protein structure directly during the reconstruction process. We have two motivations for this. First, atomic fitting is a labor-intensive step in cryo-EM post-processing, and in particular it is not clear how to perform atomic fitting for models of distributions of protein conformations learned during reconstruction with new tools such as cryoDRGN \cite{cryodrgn_nmeth}. Second, modeling the atomic structure of the 3D volume provides a strong prior over structures. In fact, in many cases the protein sequence and an approximate reference structure are known beforehand, strongly constraining the space of feasible 3D volumes. These structural priors are especially important for models of heterogeneous distributions of molecules, because they constrain the conformational dynamics to those that approximate realistic protein motions. Without such priors, it is common to observe artifacts in the motion of the protein, e.g. mass appearing and disappearing between two distinct conformations with rarely-sampled transition states.

To this end, we propose a reconstruction process based on a coarse-grained atomic model for the cryo-EM volume. The model fits parameters including atomic coordinates, to maximize the likelihood of the dataset under a generative model that maps atomic structures to cryo-EM images.
When initialized appropriately,
this approach is able to learn both homogeneous structures and heterogeneous ensembles from synthetic cryo-EM images.

\section{Background}

A cryo-EM experiment produces a dataset of $10^{4-7}$ noisy 2D projection images, each containing a unique molecule captured in a random, unknown orientation. The goal of cryo-EM reconstruction is to infer the 3D density volume $V: \mathbb{R}^3 \rightarrow \mathbb{R}$ that gave rise to the imaging dataset $X_1,...,X_N$.
As cryo-EM images are integral projections of the molecule in this imaging modality, 2D images can be related to the 3D volume by the Fourier slice theorem \cite{slicetheorem}, which states that the Fourier transform of a 2D projection is a central slice from the 3D Fourier transform of the volume. Traditional methods approximate the volume as a voxel array $\hat{V}(\bf k)$ in Fourier space \cite{Singer2020}. 

To recover the desired structure, cryo-EM reconstruction methods must jointly solve for the unknown volume $\hat{V}$ and image poses $\phi_i = (R_i,t_i)$, where $R_i \in SO(3)$ and $t_i \in \R^2$ are the 3D orientation of the molecule and in-plane image translation, respectively. Expectation maximization and simpler variants of coordinate ascent are typically employed to find a \textit{maximum a posteriori} estimate of $\hat{V}$ marginalizing over the posterior distribution of $\phi_i$'s \cite{Scheres_bayes}. 

A unique advantage of cryo-EM is its ability to image heterogeneous molecules. \textit{Heterogeneous reconstruction} algorithms aim to reconstruct a distribution of structures from the dataset. A standard approach involves extending the generative model to assume that images are generated from a mixture model of $K$ volumes $V_1,...,V_K$ \cite{scheres2005maximum,frealign}. More recently, cryoDRGN proposed a neural model to reconstruct heterogeneous ensembles of particles from cryo-EM data \cite{zhong_iclr}. CryoDRGN represents a continuous $n$-dimensional distribution over volumes as a function $\hat{V}: \R^{3+n}\rightarrow \R$ approximated by a multi-layer perceptron (MLP) with positional encoding of Cartesian coordinate inputs \cite{zhong_iclr}. To simplify reconstruction, cryoDRGN and other advanced reconstruction methods \cite{cryodrgn_nmeth, 3dva, manifoldembedding, multibody} find it sufficient to use poses $\phi$ computed using a traditional reconstruction method, and focus on the volume reconstruction.

\section{Related Work}

A large body of work has investigated continuous deformations of protein structures produced from normal mode analysis of atomic, coarse-grained, or pseudo-atomic models \cite{Bahar2005-tm, hinsen2005normal, jin2014iterative, schilbach2017structures, harastani2020hybrid}. The top N normal modes of a system where (pseudo-)atomic bonds are approximated by harmonic springs can summarize a molecule's flexing motions, which can then be used to model cryo-EM data during reconstruction \cite{hinsen2005normal, jin2014iterative, schilbach2017structures, harastani2020hybrid}. However, it is unclear how accurate the hypothetical motions that are generated from the underlying harmonic spring approximation are. In BioEM and other approaches, likelihood-based analysis of cryo-EM images or reconstructed maps based on an ensemble of atomic models (e.g. generated from molecular dynamics) has been used to investigate conformational heterogeneity \cite{cossio2013bayesian, bonomi2018simultaneous, cossio2018likelihood}.

Recently, deep learning has been used to incorporate prior beliefs from an existing dataset of atomic structures into the cryo-EM reconstruction process. Kimanius \textit{et al.} train a convolutional denoising model on PDB structures that is integrated into the cryo-EM reconstruction process in the form of updates to regularization parameters during iterative refinement \cite{kimanius2021exploiting}. They show higher resolution reconstructions of synthetic datasets but observe artifacts from the learned general model in some cases. DeepEMhancer trains a network to perform volume post-processing (e.g. high frequency sharpening and solvent/background masking) from pairs of raw and postprocessed experimental density maps, where the postprocessed maps have been refined by the fitted atomic structure \cite{deepemhancer}. 

In concurrent work, Rosenbaum \textit{et al.} propose a VAE with a similar RBF-based generative model over atomic coordinates \cite{rosenbaum2021inferring}. While we model proteins at the residue level, Rosenbaum \textit{et al.} model them at the atomic level \cite{rosenbaum2021inferring} and also jointly infer image pose. They evaluate this model on synthetic data from a molecular dynamics simulation. 

\section{Method}

\subsection{Cryofold volume model}

The central contribution of this work is a cryo-EM density model that is parameterized in terms of coordinates for a coarse-grained atomic model given a known atomic reference structure (Figure \ref{fig:architecture}A). In this coarse-grained model, each amino acid is represented by two Gaussian radial basis functions (RBFs), one representing the backbone and the second representing the sidechain. Each RBF is parameterized by $(\mu_i, \sigma_i, a_i)$, where $\mu_i$ is the position of the $i$th RBF; $a_i$ is the amplitude of the $i$th RBF; and $\sigma_i$ is the width of the $i$th RBF. We tie $a_i = a_0 Z_i$ where $a_0$ is a global learned amplitude constant, and $Z_i$ is the total number of electrons in the fragment represented by RBF $i$. Furthermore, we tie all $\sigma_i$ to the same value $\sigma$. Thus, the full RBF model has $3K + 2$ parameters, where $K$ is the total number of amino acids in the protein complex. 
The cryo-EM density can be computed as a function of these parameters as:

\begin{equation}
V(\mathbf{r}) = \sum_i (2\pi \sigma_i)^{-3/2} a_i \exp\left({\frac{-||\mathbf{r} - \mathbf{\mu_i}||^2}{2 \sigma_i^2}}\right)
\end{equation}

As described in Section 2, reconstruction is performed in the Fourier domain. This makes the choice of Gaussian RBFs convenient, as $V$ can be computed efficiently in Fourier space\footnote{Incidentally, projections of Gaussian kernels can also be computed analytically in real space, obviating the need for the Fourier slice theorem altogether. This real space formulation allows for algorithms that scale better with the number of RBFs since the RBFs are localized in real space, allowing for spatial decomposition via gridding, KD-trees, etc.}:
\begin{equation}
\hat{V}(\mathbf{k}) = \sum_i { a_i 
    e^{-2\pi i \mathbf{\mu_i\cdot k}}
    \exp
    \left(
        \frac{
            -\pi^2 \mathbf{k}^2 \sigma_i^2
        }{
            2
        }
    \right)
}
\end{equation}

To impose physical constraints on the RBF model, we add a set of fixed harmonic bond terms $\mathcal{B}$ between consecutive backbone RBF centers. Each bond term $(i, j, k, l) \in \mathcal{B}$ is specified by a pair of RBF indices $i, j$ and a bond strength $k$, which we set to 0.1. The bond length $l$ is set to 3.8 \AA, the distance between protein $C_\alpha$ backbone carbons.

We constrain the side-chain RBFs to be located close to their backbone RBF using one-sided harmonic restraints $\mathcal{C}$. These are similar to the bond terms, but only induce a loss when the distance between RBF centers exceeds the max length $l$. We set $l$ to the maximum distance between a backbone C-alpha carbon and its side chain center of mass observed in the reference structure.

For homogeneous reconstruction, we fit $\{\mu_i\}, \sigma, a_0$ directly using stochastic gradient descent (SGD). The overall loss function, given a set of $N$ images $\mathbf{X}$ with poses $\mathbf{\phi}$,  bond terms $\mathcal{B}$ and side-chain constraints $\mathcal{C}$ is:

\begin{equation}
\begin{split}
    \mathcal{L}(\mu, \sigma, a_0|\mathbf{X}) = \frac{1}{N}\sum_{(x, \phi) \in X} ||\hat{X}(\phi|\mu, \sigma, a_0) - X||^2 & + \sum_{(i, j, k, l)\in \mathcal{B}} k (||\mathbf{\mu}_i - \mathbf{\mu}_j||-l)^2\\
    & + \sum_{(i, j, k, l)\in \mathcal{C}} k~\max\left(||\mathbf{\mu}_i - \mathbf{\mu}_j||-l, 0\right)^2
\end{split}
\end{equation}

\subsection{Heterogeneous reconstruction}

\begin{figure}[t]
\centering
\includegraphics[width=\textwidth]{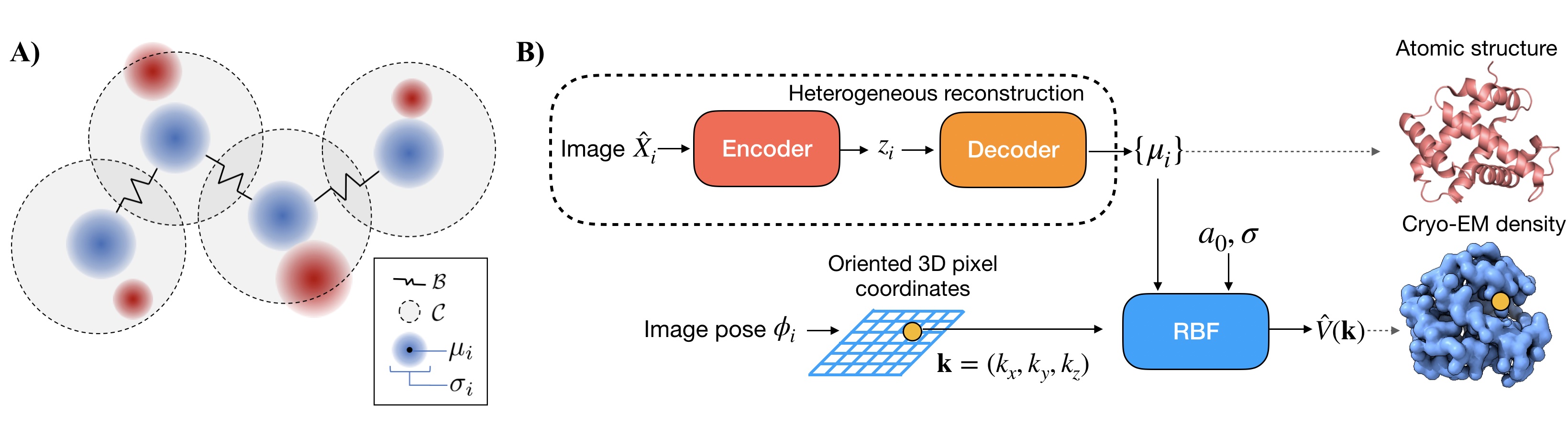}
\caption{\small{Cryofold model and architecture. \textbf{A)} The physics-inspired cryo-EM density model consists of set of Gaussian RBFs, two for each amino acid in the reference structure. One RBF represents the backbone (blue) and one represents the sidechain (red). Backbone RBFs are connected with harmonic bond terms $\mathcal{B}$, and sidechain RBFs are connected to their backbone with max-distance harmonic constraints $\mathcal{C}$.} \textbf{B)} \small{Architecture for heterogeneous reconstruction. We use a VAE to learn $z$-dependent offsets of RBF centers $\mu_i$, given an unlabelled imaging dataset of image, pose pairs $(\hat{X_i}, \phi_i)$.}}
\label{fig:architecture}
\end{figure}

For heterogeneous reconstruction, we learn a continuous latent variable model of conformational heterogeneity expressed through motion of RBF centers (Figure \ref{fig:architecture}B). Unlike standard heterogeneous cryo-EM reconstruction algorithms that use an unconstrained volume representation (e.g. voxel arrays or positionally encoded MLPs), the RBF model constrains the effect of the latent degrees of freedom to motion of the underlying atomic structure. 

An image encoder $E$ with parameters $\theta_E$ predicts a latent $z \sim E(\hat{X}|\theta_E)$; A decoder network $D$ with parameters $\theta_D$ predicts a $z$-dependent translation of the RBF centers $\mu_{het}(z) = \mu + D(z|\theta_D)$. We optimize $\mu, \sigma, a_0, \theta_D, \theta_E$ together end-to-end with SGD.

\subsection{Local minima}

A major shortcoming of this atomic reconstruction approach is the existence of many local minima of the loss function that do not approximate the true atomic coordinates. This is in contrast to voxel-based models, where SGD converges to the global minimum of the convex loss given the poses, and in contrast to large-scale generic neural models (e.g. \cite{zhong_iclr, cryodrgn_nmeth}) which typically do not suffer from problematic local minima in practice \cite{choromanska2015loss}.  We address the local minima problem primarily by initializing RBF centers $\mu_i$ from a reference structure that is a close approximation of the imaged structure (homogeneous) or some point in the distribution of structures (heterogeneous). Such reference structures are often available when studying a variant of a known structure, or heterogeneous protein dynamics.

\section{Results}

Here, we present results for reconstructing coarse-grained atomic structures from synthetic cryo-EM image data with Cryofold. We first explore the effect of reference structure initialization in homogeneous reconstruction, when initialized from either an approximate reference structure (fitting), the exact structure (cheating), or from randomly initialized locations (folding). We then turn to heterogeneous cryo-EM datasets, where we evaluate the ability of Cryofold to reconstruct continuous distributions of structures and their atomic coordinates.  Finally, we assess choices in the design of the RBF model by ablating key components in the homogeneous setting.

\textbf{Datasets.} We generate homogeneous and heterogeneous cryo-EM datasets using a  141-residue atomic model (PDB 5NI1) as the ground truth structure. To generate homogeneous data, we simulate 50,000 noisy projection images based on the cryo-EM image formation model. To model heterogeneity, we introduce a bond rotation in the backbone of 5ni1 to create a continuous 1D motion, and generate cryo-EM images sampling along the ground truth reaction coordinate. Further details on dataset generation are given in the Appendix.

\textbf{Architecture and training.} For all experiments, we train for 10 epochs using the Adam optimizer in minibatches of 8 images and a learning rate of 1e-4. We initialize $\sigma$ and $a_0$ to $3.71\AA$ and $0.1$, and perform one epoch of `warm-up' to refine these global parameters after which we reset the atomic coordinates to their initial values. For homogeneous reconstruction, we directly optimize all RBF parameters. For heterogeneous reconstruction, we use a VAE to predict the $z$-dependent offsets of the RBF centers from their reference values. Both the encoder and decoder are 3-layer MLPs of width 256 and residual connections, and a 1-dimensional latent variable. We use ground-truth poses $\phi$ for training. In real applications, the poses would be inferred from traditional cryo-EM tools \cite{cryosparc, relion3, cistem}. The model is implemented in PyTorch \cite{paszke2019pytorch}.

\subsection{Homogeneous reconstruction}

To explore the effect of reference structure initialization, we compare the reconstruction accuracy of Cryofold when initialized from the ground truth coordinates ("Exact 5NI1") and from three alternate starting configurations: 1) an approximate reference structure generated by evolving the system under a molecular dynamics simulation ("Approximate 5NI1"), 2) the ground truth structure with 6 \AA~uniform noise added to the ground truth coordinates of each $C_\alpha$ ("5NI1 + Uniform[6 \AA]"), and 3) a random initialization of each $C_\alpha$ RBF in the $64 \AA^3$ region in the center of the box ("Random"). Side-chain RBFs are initialized to their corresponding $C_\alpha$ RBF centers.

\begin{figure}[t]
\centering
\includegraphics[width=300px]{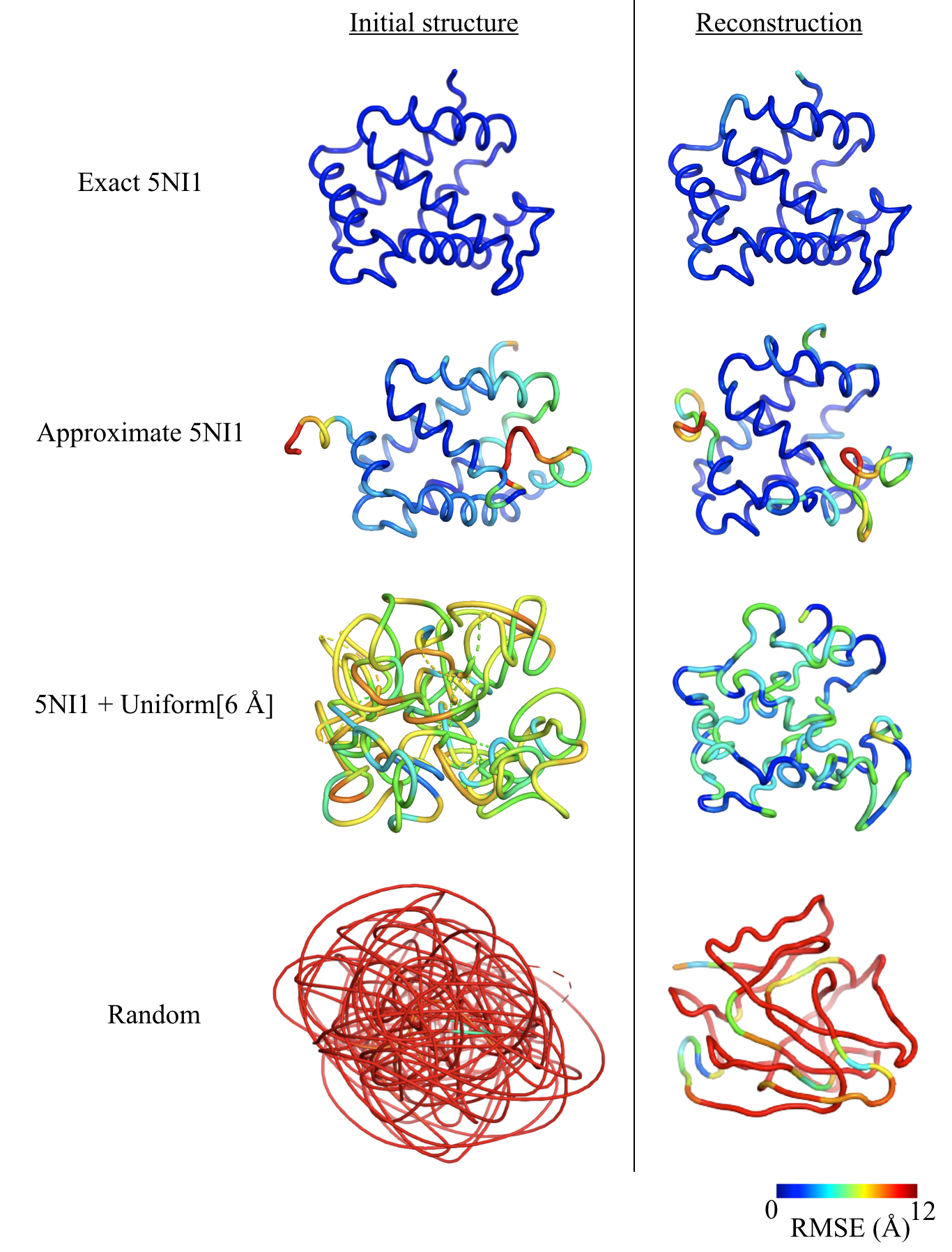}
\caption{Reconstructed atomic structures with RBF centers initialized either 1) at the exact ground truth values of 5NI1, 2) at an approximate structure generated from evolution by molecular dynamics, 3) at 5NI1 coordinates randomly perturbed by Uniform[6 \AA] noise, or 4) from random initial values. Structures are colored by $C_\alpha$ RMSE to the ground truth structure (top left).}
\label{fig:initialization}
\end{figure}

We report the root-mean-square-error (RMSE) of model backbone coordinates to the $C_\alpha$ of the true structure, the percent of $C_\alpha$ backbone atoms predicted within 3 \AA\ of the true structure, and the normalized mean-square-error (NMSE) of the reconstructed volume to the true structure (Table \ref{tab:homo_init}). We find that our atomic model is quite sensitive to initialization. While the model performs adequately when initialized at nearby reference structures, it makes some mistakes due to local minima, and performs much better when initialized with the ground truth coordinates. When initialized from random coordinates, SGD is unable to recover the ground truth atomic coordinates whatsoever. Instances of atomic local minima include `mismatched buttoning' of the amino acid backbone where multiple backbone RBFs are trapped in the same location, as well as incorrect tracing of the protein backbone through the volume (Figure \ref{fig:initialization}). 

\begin{table}[th]
\begin{center}
\begin{tabular}{ l | r | r r r}
\toprule
{\bf Initialization} & \bf Initial RMSE (\AA) & {\bf $C_\alpha$ RMSE (\AA)} & {\bf \% within 3\AA} & {\bf Volume NMSE} \\
\midrule
Exact 5NI1 & 0.00 & 0.843 & 99.29\% & 0.28 \\ 
Approximate 5NI1 & 5.15 & 3.81 & 77.30\% & 0.32 \\ 
5NI1 + Uniform[6 \AA] & 6.50 & 3.19 & 53.19\% & 0.35 \\ 
Random & 34.87 & 17.16 & 2.12\% & 0.43  \\ 
\bottomrule
\end{tabular}
\end{center}
\caption{\small Comparison of different choices of initial atomic coordinates when training the model.} 
\label{tab:homo_init}
\end{table}

\subsection{Heterogeneous reconstruction}

The main motivation behind proposing this model is to provide inductive bias from known atomic structures when reconstructing heterogeneous protein conformations. 
As opposed to the previous section where we reconstruct a single static structure, here we attempt to reconstruct a continuous manifold of structures from a heterogeneous cryo-EM dataset. 

We consider a synthetic dataset of noisy projection images, where the underlying protein structure possesses a 1D continuous motion of the 5NI1 protein. A bond in the center of the protein is rotated, leading to a large-scale global conformational change (Figure \ref{fig:het2_ground_truth}). In our experiments, we initialize the RBF model to the structure at one end of the reaction coordinate, and train a heterogeneous RBF model with a 1-dimensional latent variable on the imaging dataset with ground truth poses. 

\begin{figure}[th]
\centering
\includegraphics[width=\textwidth]{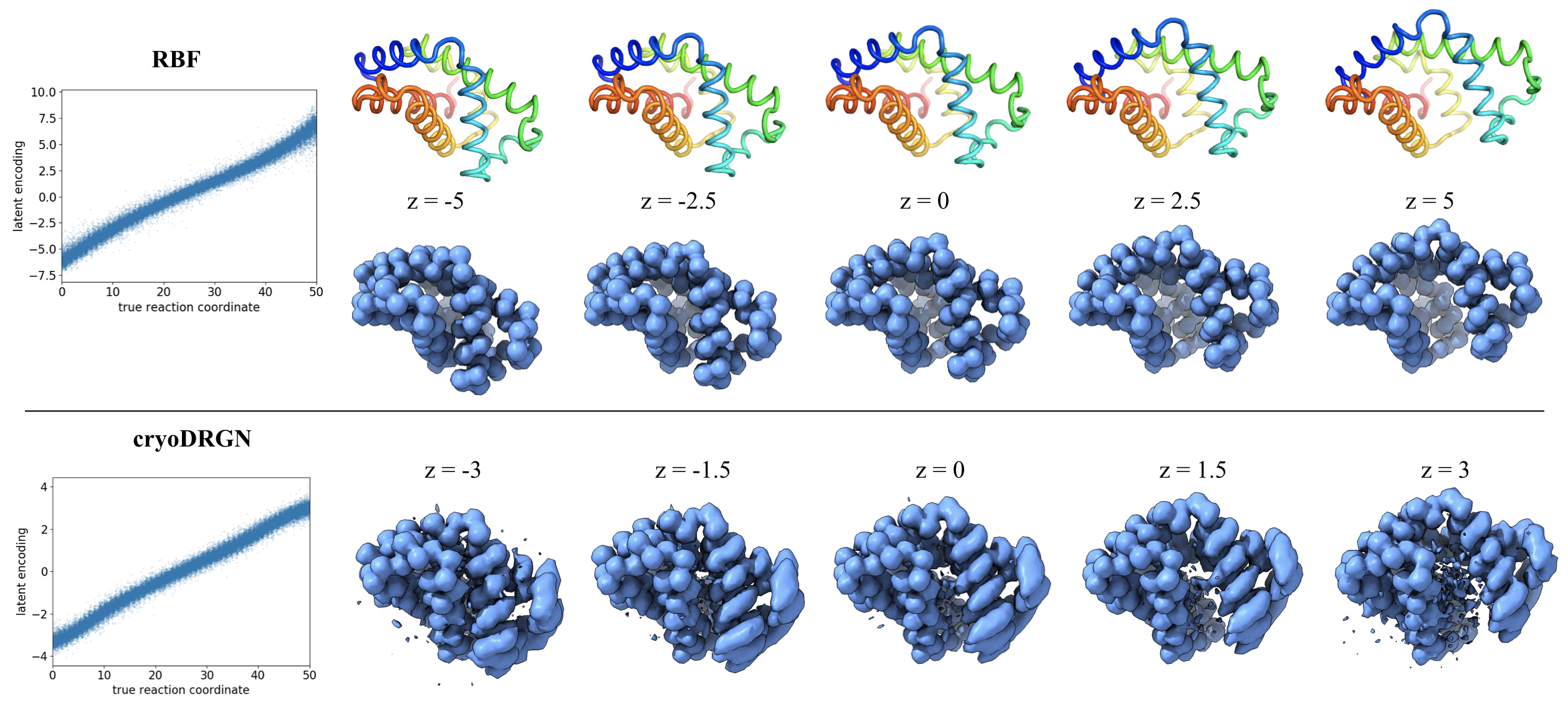}
\caption{\small Heterogeneous reconstruction results of an unlabeled dataset containing a uniform distribution of images across the ground truth reaction coordinate. Predicted 1D latent encoding $z$ plotted against the ground truth reaction coordinate (left), and reconstructed structures at the specified values of $z$. Cryofold directly reconstructs atomic coordinates (top). The unconstrained neural volume representation (cryoDRGN) contains noise and blurring of moving atoms in the reconstructed volumes (bottom) and does not produce atomic coordinates.}
\label{fig:het2_results}
\end{figure}

The RBF model is able to correctly reconstruct the full distribution of 3D structures containing this large conformational change. The latent encodings of the images are well correlated with the true reaction coordinate (Figure \ref{fig:het2_results}, left), and the RBF atomic coordinates from traversing the latent space nearly exactly reconstruct the underlying protein motion (Figure \ref{fig:het2_results}, right).

As a baseline, we perform heterogeneous reconstruction with cryoDRGN \cite{cryodrgn_nmeth}, which learns an unconstrained neural representation of cryo-EM volumes. Similarly, we provide the ground truth poses and train a 1-D latent variable model with identical encoder architecture and training settings. While the latent space is well correlated with the ground truth motion similar to the RBF model, the reconstructed volumes from cryoDRGN contain noise and blurring artifacts in the mobile region whereas Cryofold volumes are regularized using structural priors from the underlying atomic model (Figure \ref{fig:het2_results}, bottom). 

\begin{figure}[t]
\centering
\includegraphics[width=195pt]{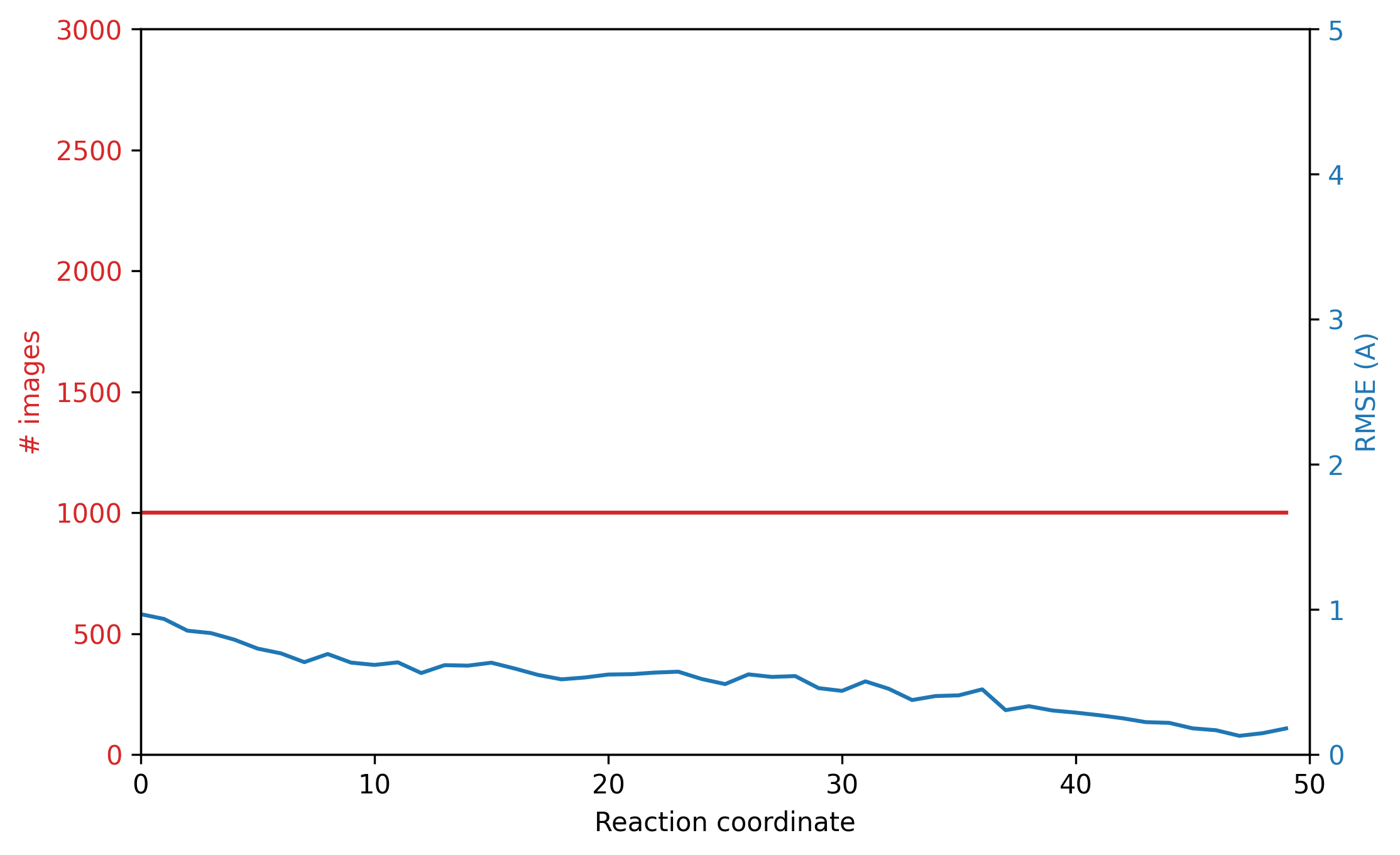}
\includegraphics[width=195pt]{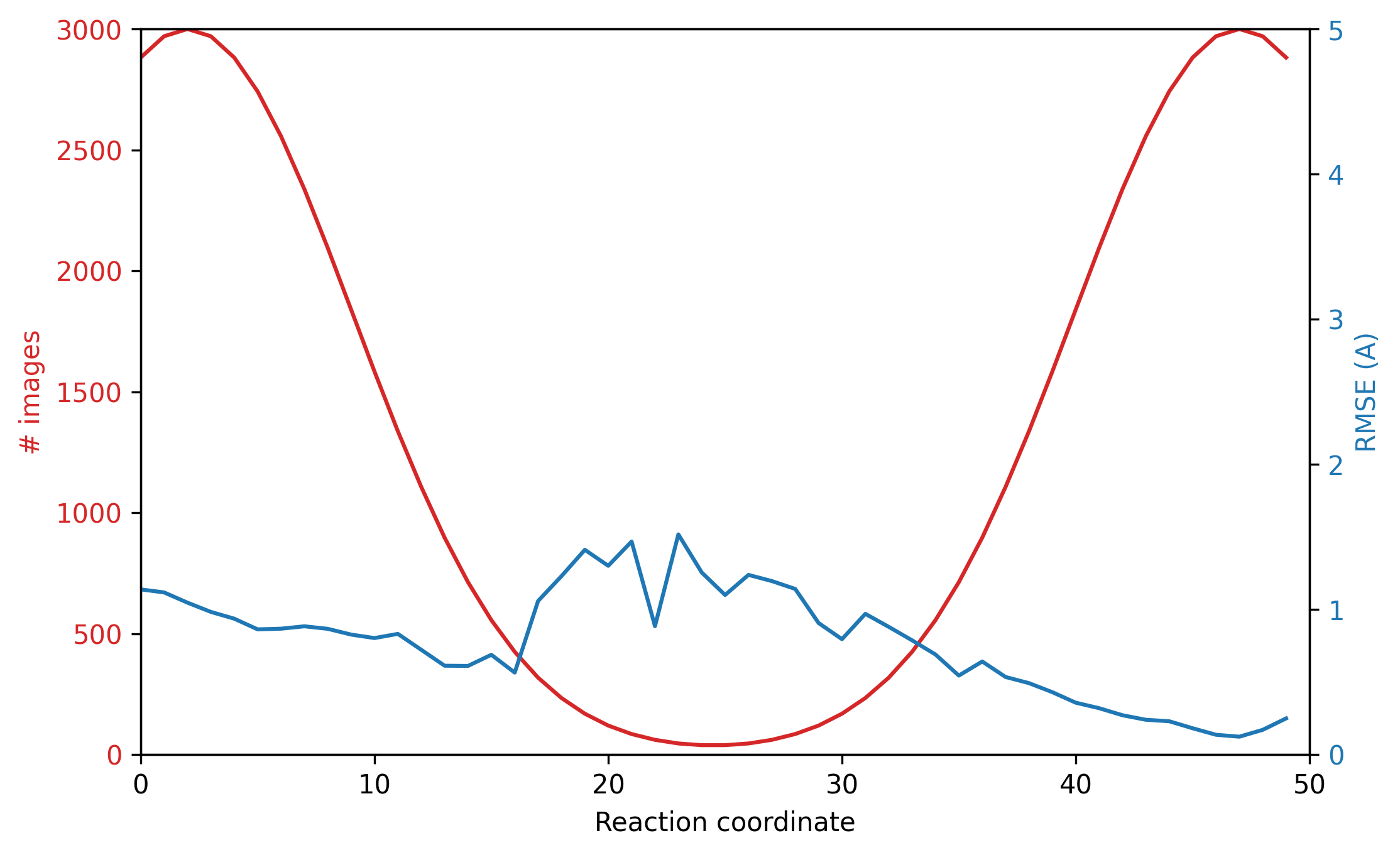}
\includegraphics[width=195pt]{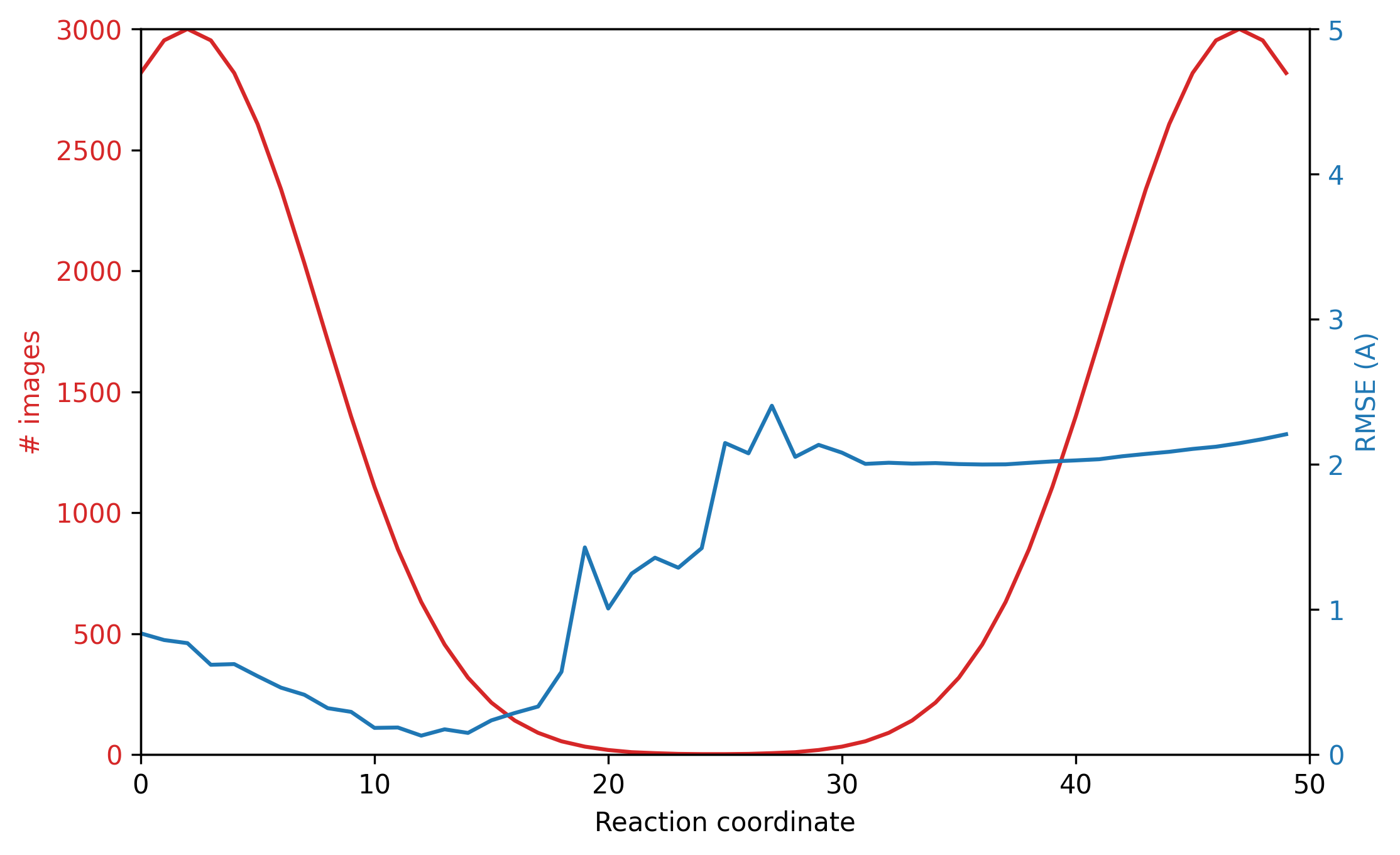}
\includegraphics[width=195pt]{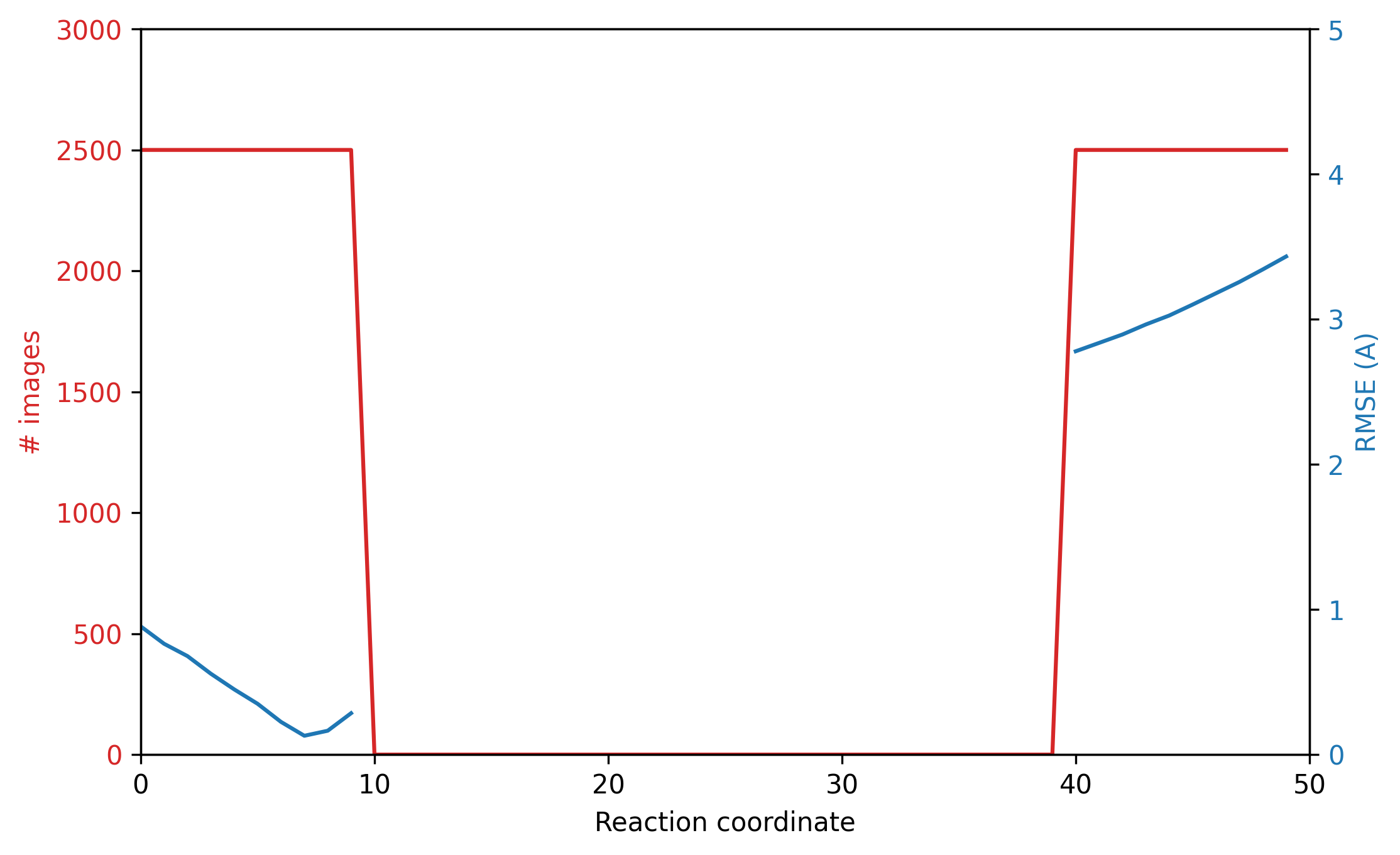}
\caption{\small $C_\alpha$ RMSE (blue) for heterogeneous reconstruction of a large synthetic conformational change of 5ni1, with different distributions of images (red) across the reaction coordinate. The reference structure used for initialization is the ground truth atomic structure at reaction coordinate 0.}
\label{fig:het2_distribution}
\end{figure}

We also measure reconstruction accuracy across the reaction coordinate for four distributions of images across the reaction coordinate: a uniform distribution, two non-uniform distributions corresponding to an energy barrier of different heights between the end states; and two discrete clusters with no images in the middle (Figure \ref{fig:het2_distribution}). For each value of the reaction coordinate, we approximate the atomic coordinates using the median latent from the images at that coordinate, and measure its $C_\alpha$ RMSE with the ground truth. We see that when there is even a small probability mass across the reaction coordinate (Figure \ref{fig:het2_distribution}, top row), the atomic model learns the full distribution of conformations with high accuracy. However, if transition states are nearly or completely unobserved in the image distribution (Figure \ref{fig:het2_distribution}, bottom row), reconstruction accuracy is poor.


\subsection{Model ablations}

Using the homogeneous dataset and an initialization from the approximate 5NI1 structure, we explore various choices in design of the RBF model by ablating the side chain RBF, bond constraints, and the cryo-EM supervision (Table \ref{tab:homo_ablations}). We find that removing the sidechain RBFs and modeling each amino acid as a single RBF slightly degrades quality, whereas removing the internal bond terms leads to a dramatic degradation. Atomic accuracy is substantially worse but not completely degraded even when ignoring the cryo-EM images due to the initialization; however, as expected there is low overlap with the true volume in this case. We expect the volume NMSE to be higher than unconstrained approaches (e.g. cryoDRGN), as the coarse-grained RBF model does not exactly match the underlying all-atom generative model even with correct coordinates (Figure \ref{fig:exact_initialization_volumes}). 

\begin{table}[h!]
\begin{center}
\begin{tabular}{ l | r r r }
\toprule
{\bf Model} & {\bf $C_\alpha$ RMSE (\AA)} & {\bf \% within 3\AA} & {\bf Volume NMSE} \\
\midrule
Full Model & \textbf{3.81} & \textbf{77.30\%} & \textbf{0.32}  \\
{\it No Sidechain RBFs} & 4.10 & 70.21\% & 0.49  \\
\it No Bonds & 10.42& 37.59\% & 0.36  \\
\it No Cryo-EM Loss & 5.15 & 74.47\% & 1.82 \\ 
\midrule
CryoDRGN & {N/A} & {N/A} & 0.08 \\
\bottomrule
\end{tabular}
\end{center}
\caption{\small Ablations of model components. We remove the sidechain RBFs, the bond terms between RBFs, and the cryo-EM reconstruction loss; each degrades the quality of reconstruction. }
\label{tab:homo_ablations}
\end{table}

\section{Discussion}

This work presents Cryofold, a simple framework for incorporating prior information provided from an atomic model into cryo-EM reconstruction. By constraining the generative model to act on a (coarse-grained) atomic model, we constrain the model output to a submanifold of coordinate positions, which both provides strong regularization to reduce artifacts seen in unconstrained methods and directly yields interpretable atomic coordinates. Our experiments suggest that such models are a promising direction for incorporating structural priors into cryo-EM reconstruction, especially for heterogeneous structures. However the experimental validation is preliminary: we worked entirely with \textit{synthetic} cryo-EM datasets of a \textit{small} protein using \textit{exact poses}. Follow-up work is required to understand how these techniques behave with real cryo-EM images and volumes, using realistic protein complexes of interest, and using approximate poses generated by existing tools. For larger protein complexes, follow-up work will investigate whether a coarser model granularity may be more appropriate, especially when modeling large-scale heterogeneous dynamics.

Results on homogeneous datasets suggest that local minima in optimization space are a major problem for this class of methods if coordinates are not initialized near the ground truth structure. These local minima problems could potentially be ameliorated with a combination of improved modeling of structural priors such as additional bonded terms and steric interactions, and optimization methods such as Hamiltonian dynamics or Markov Chain Monte Carlo that can escape local minima. There is a rich literature on these topics in the domain of molecular dynamics simulation which could carry over to cryo-EM reconstruction \cite{bernardi2015enhanced,yang2019enhanced}. 

However, results on heterogeneous datasets suggest that we can correctly learn distributions with large continuous conformational changes when initialized from some structure in the distribution. Even structural changes that are too large to be modeled correctly when only the endpoint structures are observed (as in homogeneous reconstructions) were successfully reconstructed from heterogeneous datasets. Additional work is required to more fully characterize exactly when neural models of heterogeneous structures converge to the correct distribution.

It should be noted, however, that prior information \textit{biases} the model output. In particular, this approach will not be able to model other sources of image heterogeneity that can not be expressed as motion of the RBFs (e.g. compositional variation) whose presence are not known \textit{a priori}. Importantly, the separation of atomic modelling from volume reconstruction in existing cryo-EM pipelines provides a crucial mechanism for validation of the volume reconstruction accuracy. This and related approaches that incorporate atomic priors should be carefully employed only in settings where prior beliefs are strongly held, and we believe validation of these methods is an open problem. Nevertheless, future extensions of this method that can flexibly integrate biophysical simulation, learning on existing structures, and experimental single particle cryo-EM data is a promising direction for improved protein structure determination.

\section{Acknowledgements}

We thank the MIT-IBM Satori team for GPU computing resources and support. This work was funded by the National Science Foundation Graduate Research Fellowship Program to E.D.Z., NIH grant R01-GM081871 to B.B., NIH grant R00-AG050749 to J.H.D., and a grant from the MIT J-Clinic for Machine Learning and Health to J.H.D. and B.B.

\bibliographystyle{unsrt}
\bibliography{references}

\begin{thebibliography}{10}

\bibitem{Nogales:2016bv}
Eva Nogales.
\newblock {The development of cryo-EM into a mainstream structural biology
  technique}.
\newblock {\em Nature Methods}, 13(1):24--27, jan 2016.

\bibitem{cryodrgn_nmeth}
Ellen~D Zhong, Tristan Bepler, Bonnie Berger, and Joseph~H Davis.
\newblock {CryoDRGN: reconstruction of heterogeneous cryo-EM structures using
  neural networks}.
\newblock {\em Nature methods}, 18(2):176--185, 2021.

\bibitem{3dva}
Ali Punjani and David~J Fleet.
\newblock 3d variability analysis: Resolving continuous flexibility and
  discrete heterogeneity from single particle cryo-em.
\newblock {\em Journal of Structural Biology}, 213(2):107702, 2021.

\bibitem{multibody}
Takanori Nakane, Dari Kimanius, Erik Lindahl, and Sjors~Hw Scheres.
\newblock {Characterisation of molecular motions in cryo-EM single-particle
  data by multi-body refinement in RELION.}
\newblock {\em eLife}, 7:e36861, June 2018.

\bibitem{topaz}
T~Bepler, A~Morin, M~Rapp, J~Brasch, and L~Shapiro.
\newblock {Positive-unlabeled convolutional neural networks for particle
  picking in cryo-electron micrographs}.
\newblock {\em Nature Methods}, 2019.

\bibitem{cryosparc}
Ali Punjani, John~L Rubinstein, David~J Fleet, and Marcus~A Brubaker.
\newblock {cryoSPARC: algorithms for rapid unsupervised cryo-EM structure
  determination.}
\newblock {\em Nature Methods}, 14(3):290--296, March 2017.

\bibitem{relion3}
Jasenko Zivanov, Takanori Nakane, Bj{\"{o}}rn~O. Forsberg, Dari Kimanius,
  Wim~J.H. Hagen, Erik Lindahl, and Sjors~H.W. Scheres.
\newblock {New tools for automated high-resolution cryo-EM structure
  determination in RELION-3}.
\newblock {\em eLife}, 7, nov 2018.

\bibitem{cistem}
Timothy Grant, Alexis Rohou, and Nikolaus Grigorieff.
\newblock {cisTEM, user-friendly software for single-particle image
  processing}.
\newblock {\em eLife}, 7:e14874, mar 2018.

\bibitem{eman2}
Guang Tang, Liwei Peng, Philip~R Baldwin, Deepinder~S Mann, Wen Jiang, Ian
  Rees, and Steven~J Ludtke.
\newblock {EMAN2: an extensible image processing suite for electron
  microscopy.}
\newblock {\em Journal of structural biology}, 157(1):38--46, January 2007.

\bibitem{coot}
P~Emsley, B~Lohkamp, W~G Scott, and K~Cowtan.
\newblock {Features and development of Coot}.
\newblock {\em Acta Crystallographica Section D: Biological Crystallography},
  66(4):486--501, April 2010.

\bibitem{phenix}
Dorothee Liebschner, Pavel~V. Afonine, Matthew~L. Baker, G{\'{a}}bor Bunkoczi,
  Vincent~B. Chen, Tristan~I. Croll, Bradley Hintze, Li~Wei Hung, Swati Jain,
  Airlie~J. McCoy, Nigel~W. Moriarty, Robert~D. Oeffner, Billy~K. Poon,
  Michael~G. Prisant, Randy~J. Read, Jane~S. Richardson, David~C. Richardson,
  Massimo~D. Sammito, Oleg~V. Sobolev, Duncan~H. Stockwell, Thomas~C.
  Terwilliger, Alexandre~G. Urzhumtsev, Lizbeth~L. Videau, Christopher~J.
  Williams, and Paul~D. Adams.
\newblock {Macromolecular structure determination using X-rays, neutrons and
  electrons: Recent developments in Phenix}.
\newblock {\em Acta Crystallographica Section D: Structural Biology},
  75(10):861--877, oct 2019.

\bibitem{mdff}
Leonardo~G Trabuco, Elizabeth Villa, Kakoli Mitra, Joachim Frank, and Klaus
  Schulten.
\newblock {Flexible fitting of atomic structures into electron microscopy maps
  using molecular dynamics.}
\newblock {\em Structure (London, England : 1993)}, 16(5):673--683, May 2008.

\bibitem{Kihara:2020cf}
Daisuke Kihara, Genki Terashi, and Sai Raghavendra Maddhuri~Venkata
  Subramaniya.
\newblock {De Novo Computational Protein Tertiary Structure Modeling Pipeline
  for Cryo-EM Maps of Intermediate Resolution}.
\newblock {\em Biophysical Journal}, 118(3):292a, feb 2020.

\bibitem{Si:2020eg}
Dong Si, Spencer~A Moritz, Jonas Pfab, Jie Hou, Renzhi Cao, Liguo Wang, Tianqi
  Wu, and Jianlin Cheng.
\newblock {Deep Learning to Predict Protein Backbone Structure from
  High-Resolution Cryo-EM Density Maps}.
\newblock {\em Scientific Reports}, 10(1):1--22, mar 2020.

\bibitem{isolde}
Tristan~Ian Croll.
\newblock {ISOLDE: A physically realistic environment for model building into
  low-resolution electron-density maps}.
\newblock {\em Acta Crystallographica Section D: Structural Biology}, 2018.

\bibitem{Singer2020}
Amit Singer and Fred~J. Sigworth.
\newblock {Computational Methods for Single-Particle Electron Cryomicroscopy}.
\newblock {\em Annual Review of Biomedical Data Science}, 3(1):163--190, jul
  2020.

\bibitem{walls2020structure}
Alexandra~C Walls, Young-Jun Park, M~Alejandra Tortorici, Abigail Wall,
  Andrew~T McGuire, and David Veesler.
\newblock Structure, function, and antigenicity of the sars-cov-2 spike
  glycoprotein.
\newblock {\em Cell}, 2020.

\bibitem{slicetheorem}
Ronald~N Bracewell.
\newblock Strip integration in radio astronomy.
\newblock {\em Australian Journal of Physics}, 9(2):198--217, 1956.

\bibitem{Scheres_bayes}
Sjors H~W Scheres.
\newblock {A Bayesian view on cryo-EM structure determination.}
\newblock {\em Journal of Molecular Biology}, 415(2):406--418, January 2012.

\bibitem{scheres2005maximum}
Sjors~HW Scheres, Mikel Valle, Rafael Nu{\~n}ez, Carlos~OS Sorzano, Roberto
  Marabini, Gabor~T Herman, and Jose-Maria Carazo.
\newblock Maximum-likelihood multi-reference refinement for electron microscopy
  images.
\newblock {\em Journal of Molecular Biology}, 348(1):139--149, 2005.

\bibitem{frealign}
{Lyumkis, Dmitry}, {Brilot, Axel F}, {Theobald, Douglas L}, and {Grigorieff,
  Nikolaus}.
\newblock {Likelihood-based classification of cryo-EM images using FREALIGN.}
\newblock {\em Journal of Structural Biology}, 183(3):377--388, September 2013.

\bibitem{zhong_iclr}
Ellen~D Zhong, Tristan Bepler, Joseph~H Davis, and Bonnie Berger.
\newblock {Reconstructing continuous distributions of 3D protein structure from
  cryo-EM images}.
\newblock {\em ICLR}, 2020.

\bibitem{manifoldembedding}
Joachim Frank and Abbas Ourmazd.
\newblock {Continuous changes in structure mapped by manifold embedding of
  single-particle data in cryo-EM.}
\newblock {\em Methods}, 100:61--67, May 2016.

\bibitem{Bahar2005-tm}
Ivet Bahar and A~J Rader.
\newblock Coarse-grained normal mode analysis in structural biology.
\newblock {\em Curr. Opin. Struct. Biol.}, 15(5):586--592, October 2005.

\bibitem{hinsen2005normal}
Konrad Hinsen, Nathalie Reuter, Jorge Navaza, David~L Stokes, and Jean-Jacques
  Lacap{\`e}re.
\newblock Normal mode-based fitting of atomic structure into electron density
  maps: application to sarcoplasmic reticulum ca-atpase.
\newblock {\em Biophysical journal}, 88(2):818--827, 2005.

\bibitem{jin2014iterative}
Qiyu Jin, Carlos Oscar~S Sorzano, Jos{\'e}~Miguel De~La Rosa-Trev{\'\i}n,
  Jos{\'e}~Rom{\'a}n Bilbao-Castro, Rafael N{\'u}{\~n}ez-Ram{\'\i}rez, Oscar
  Llorca, Florence Tama, and Slavica Joni{\'c}.
\newblock Iterative elastic 3d-to-2d alignment method using normal modes for
  studying structural dynamics of large macromolecular complexes.
\newblock {\em Structure}, 22(3):496--506, 2014.

\bibitem{schilbach2017structures}
Sandra Schilbach, Merle Hantsche, Dmitry Tegunov, Christian Dienemann, Cristoph
  Wigge, Henning Urlaub, and Patrick Cramer.
\newblock Structures of transcription pre-initiation complex with tfiih and
  mediator.
\newblock {\em Nature}, 551(7679):204--209, 2017.

\bibitem{harastani2020hybrid}
Mohamad Harastani, Carlos Oscar~S Sorzano, and Slavica Joni{\'c}.
\newblock Hybrid electron microscopy normal mode analysis with scipion.
\newblock {\em Protein Science}, 29(1):223--236, 2020.

\bibitem{cossio2013bayesian}
Pilar Cossio and Gerhard Hummer.
\newblock Bayesian analysis of individual electron microscopy images: Towards
  structures of dynamic and heterogeneous biomolecular assemblies.
\newblock {\em Journal of structural biology}, 184(3):427--437, 2013.

\bibitem{bonomi2018simultaneous}
Massimiliano Bonomi, Riccardo Pellarin, and Michele Vendruscolo.
\newblock Simultaneous determination of protein structure and dynamics using
  cryo-electron microscopy.
\newblock {\em Biophysical journal}, 114(7):1604--1613, 2018.

\bibitem{cossio2018likelihood}
Pilar Cossio and Gerhard Hummer.
\newblock Likelihood-based structural analysis of electron microscopy images.
\newblock {\em Current opinion in structural biology}, 49:162--168, 2018.

\bibitem{kimanius2021exploiting}
Dari Kimanius, Gustav Zickert, Takanori Nakane, Jonas Adler, Sebastian Lunz,
  C-B Sch{\"o}nlieb, Ozan {\"O}ktem, and Sjors~HW Scheres.
\newblock Exploiting prior knowledge about biological macromolecules in cryo-em
  structure determination.
\newblock {\em IUCrJ}, 8(1), 2021.

\bibitem{deepemhancer}
R~S{\'a}nchez-Garc{\'\i}a, J~Gomez-Blanco, A~Cuervo, J~M Carazo, COS Sorzano,
  and J~Vargas.
\newblock {DeepEMhancer: a deep learning solution for cryo-EM volume
  post-processing}.
\newblock {\em bioRxiv}, 2020.

\bibitem{rosenbaum2021inferring}
Dan Rosenbaum, Marta Garnelo, Michal Zielinski, Charlie Beattie, Ellen Clancy,
  Andrea Huber, Pushmeet Kohli, Andrew~W. Senior, John Jumper, Carl Doersch,
  S.~M.~Ali Eslami, Olaf Ronneberger, and Jonas Adler.
\newblock Inferring a continuous distribution of atom coordinates from cryo-em
  images using vaes, 2021.

\bibitem{choromanska2015loss}
Anna Choromanska, Mikael Henaff, Michael Mathieu, Gérard~Ben Arous, and Yann
  LeCun.
\newblock The loss surfaces of multilayer networks.
\newblock 2015.

\bibitem{paszke2019pytorch}
Adam Paszke, Sam Gross, Francisco Massa, Adam Lerer, James Bradbury, Gregory
  Chanan, Trevor Killeen, Zeming Lin, Natalia Gimelshein, Luca Antiga, Alban
  Desmaison, Andreas Köpf, Edward Yang, Zach DeVito, Martin Raison, Alykhan
  Tejani, Sasank Chilamkurthy, Benoit Steiner, Lu~Fang, Junjie Bai, and Soumith
  Chintala.
\newblock Pytorch: An imperative style, high-performance deep learning library.
\newblock {\em NeurIPS}, 2019.

\bibitem{bernardi2015enhanced}
Rafael~C Bernardi, Marcelo~CR Melo, and Klaus Schulten.
\newblock Enhanced sampling techniques in molecular dynamics simulations of
  biological systems.
\newblock {\em Biochimica et Biophysica Acta (BBA)-General Subjects},
  1850(5):872--877, 2015.

\bibitem{yang2019enhanced}
Yi~Isaac Yang, Qiang Shao, Jun Zhang, Lijiang Yang, and Yi~Qin Gao.
\newblock Enhanced sampling in molecular dynamics.
\newblock {\em The Journal of chemical physics}, 151(7):070902, 2019.

\bibitem{chimera}
Eric~F Pettersen, Thomas~D Goddard, Conrad~C Huang, Gregory~S Couch, Daniel~M
  Greenblatt, Elaine~C Meng, and Thomas~E Ferrin.
\newblock {UCSF Chimera: A visualization system for exploratory research and
  analysis}.
\newblock {\em Journal of Computational Chemistry}, 25(13):1605--1612, oct
  2004.

\bibitem{pygo}
Ellen~D Zhong and Michael~R Shirts.
\newblock {Thermodynamics of Coupled Protein Adsorption and Stability Using
  Hybrid Monte Carlo Simulations}.
\newblock {\em Langmuir}, 30(17):4952--4961, 2014.

\end{thebibliography}

\vfill
\pagebreak
\normalsize

\appendix 

\newcommand{\beginsupplement}{%
        \setcounter{table}{0}
        \renewcommand{\thetable}{S\arabic{table}}%
        \setcounter{figure}{0}
        \renewcommand{\thefigure}{S\arabic{figure}}%
     }

\section{Appendix}
\beginsupplement

\subsection{Dataset generation}

\begin{figure}[th]
\centering
\includegraphics[width=200px]{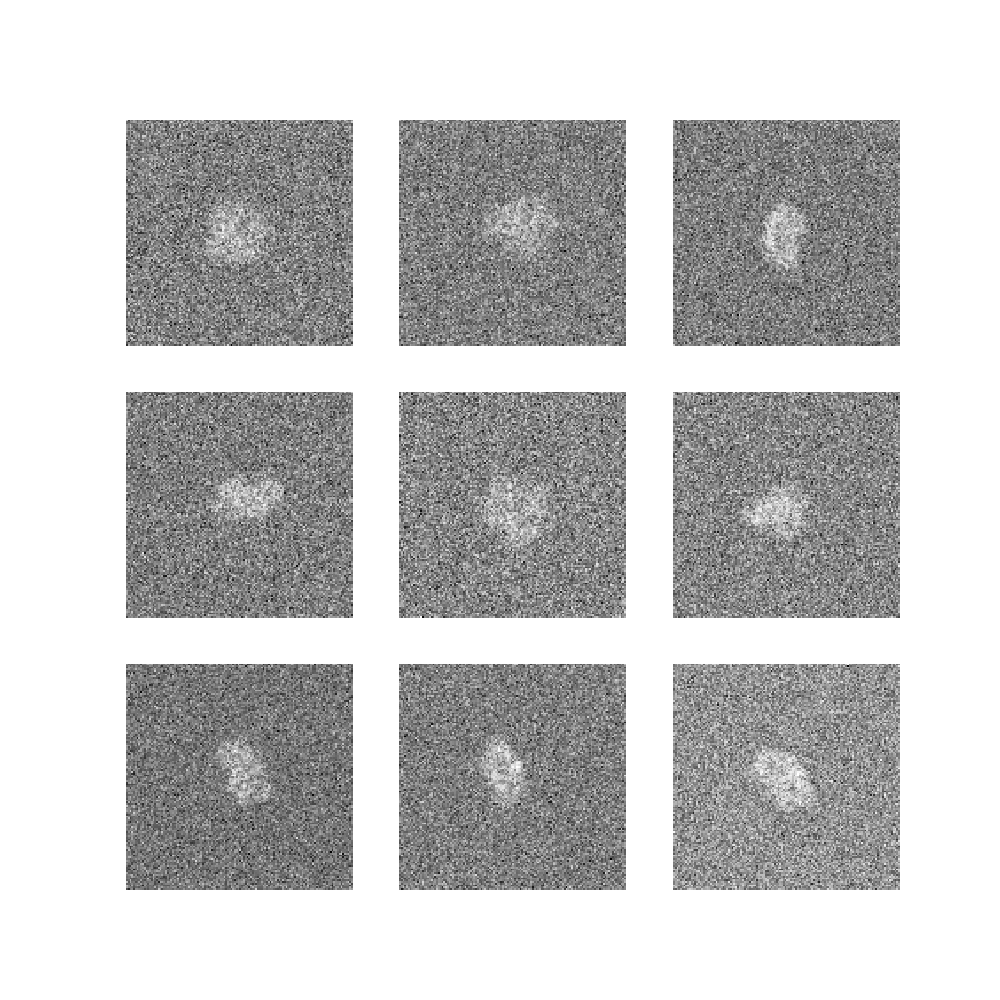}
\caption{\small{Example projection images.}}
\label{fig:particles}
\end{figure}

Synthetic cryo-EM datasets were generated from an atomic model of PDB 5NI1 according to the image formation model as follows: starting from the deposited atomic model of 5NI1, the 141-residue A-chain subunit of the complex was extracted. A cryo-EM volume was generated with the `molmap` command in Chimera \cite{chimera} at 3 \AA\  resolution and grid spacing of 1 \AA. The volume was zero-padded to a cubic dimension of $128^3$. 50,000 projection images were generated at random orientations of the volume uniformly from SO(3). Images were then translated in-plane by $t$ uniformly sampled from $[-10,10]^2$ pixels. We omit the application of the CTF for simplicity in this synthetic dataset. Gaussian noise was added leading to a signal to noise ratio (SNR) of 0.1, where we define the whole image as the signal (Figure \ref{fig:particles}). As real cryo-EM datasets have variable resolution and thus resolvability of the atomic structure, we investigate the effect of volume resolution and the included atoms in generating the dataset's ground truth volume/images (Table \ref{tab:homo_generative}). We find that our atomic modeling is quite robust to the resolution and which atoms we model in the synthetic data, which suggests that it may transfer well to real cryo-EM images.

To generate the "Approximate 5NI1" reference structure, we evolve a coarse-grained $C_\alpha$-only model of the 5NI1 A chain under a short Hamiltonian Monte Carlo simulation near the system's melting temperature. We use the implementation from \cite{pygo}, and run 10k Monte Carlo moves where each move consists of 2.25 ps of molecular dynamics at 320 K. The final frame of the trajectory is used as the approximate reference structure.

\begin{figure}[t]
\centering
\includegraphics[width=\textwidth]{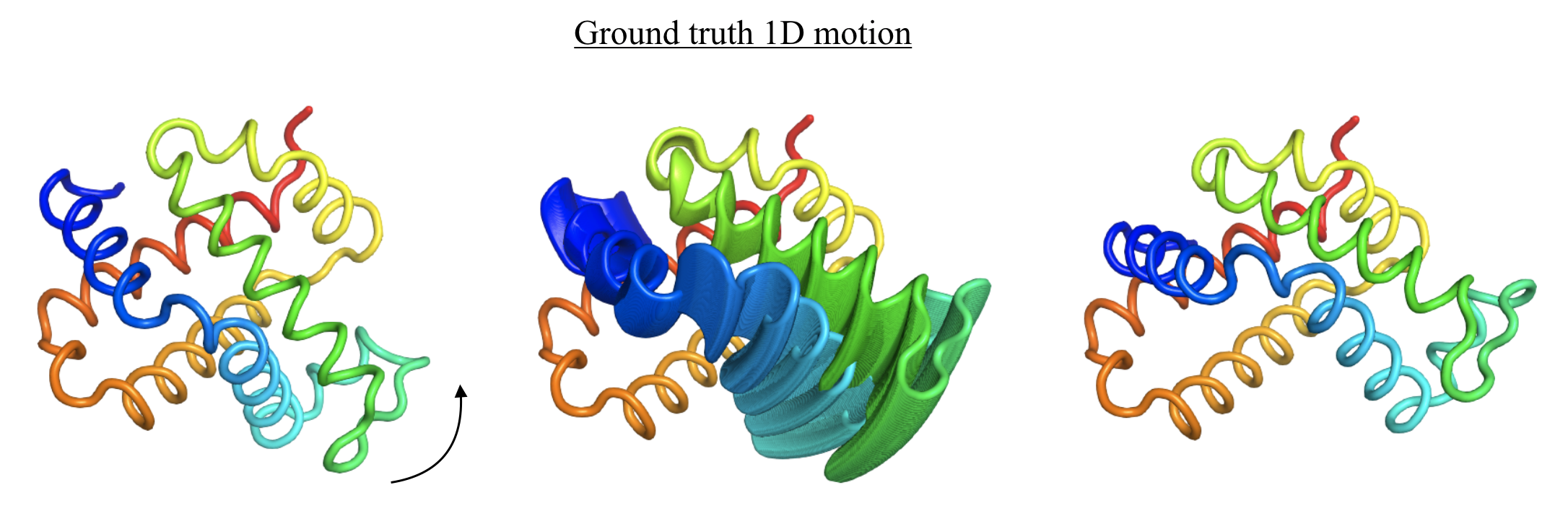}
\caption{\small{Ground truth structures of the heterogeneous datasets simulating a 1D continuous motion transitioning from left (5NI1) to right. All generated structures are shown in the center.}}
\label{fig:het2_ground_truth}
\end{figure}

To create synthetic heterogeneous datasets, a dihedral angle in the backbone of the atomic model was rotated through 0.25 radians with a structure generated every 0.05 radians along the motion (Figure \ref{fig:het2_ground_truth}). The resulting 50 structures are used as the ground truth structures for approximating the continuous motion. For simplicity of heterogeneous dataset generation, we use the 5NI1 atomic models with $C_\alpha$ backbone atoms only. We generate multiple datasets of 50k total images following the homogeneous dataset image formation process, each with a different distribution of images along the reaction coordinate as shown in Figure \ref{fig:het2_distribution}.  

\begin{figure}[t]
\centering
\includegraphics[width=300px]{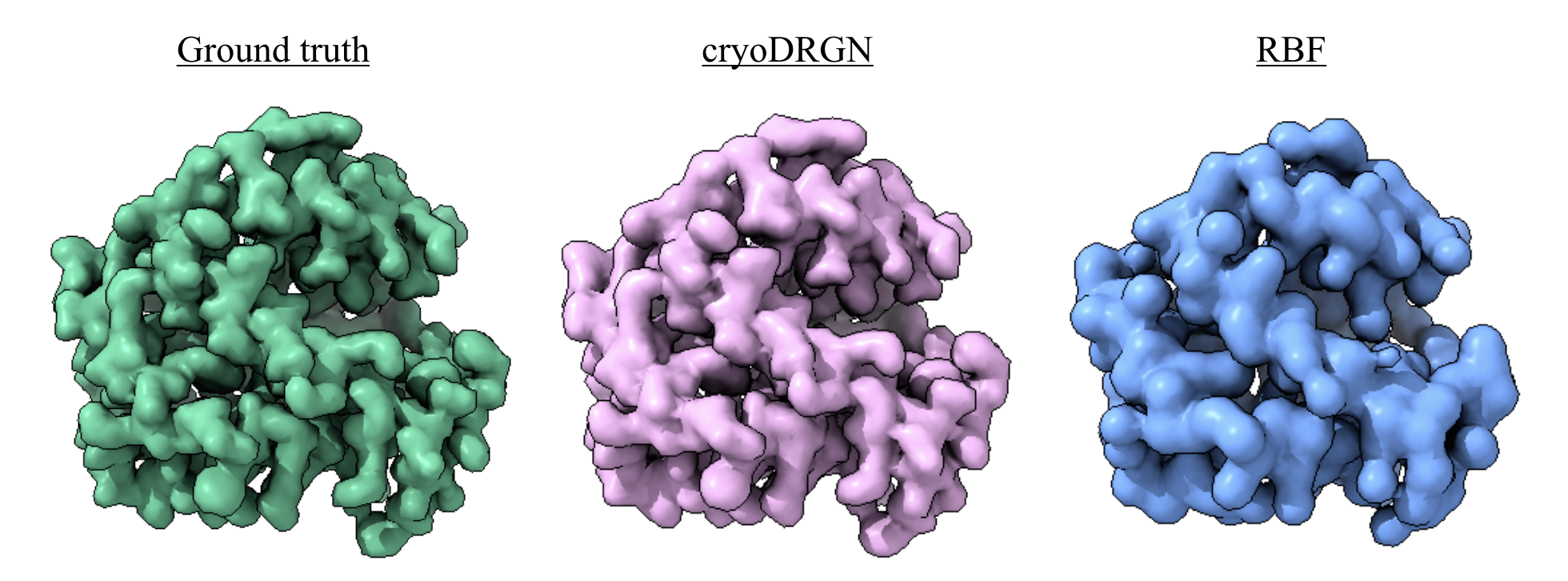}
\caption{Ground truth and reconstructed volumes with a neural network representation of density (cyroDRGN) and with our RBF model initialized from exact coordinates. The RBF volume reconstruction is somewhat worse than that of cryoDRGN because even with exact coordinates, it cannot match the all-atom generative model.}
\label{fig:exact_initialization_volumes}
\end{figure}

%

\begin{table}[h!]
\begin{center}
\begin{tabular}{ l  c | c c c }
\toprule
{\bf Modeled Atoms} & {\bf Resolution} & {\bf $C_\alpha$ RMSE (\AA)} & {\bf \% within 3\AA} \\
\midrule
C$\alpha$ & 3\AA & 3.21 & 80.8\% & \\
C$\alpha$ \& C$\beta$ & 3\AA & 3.29 & 77.3\% & \\
All Atoms & 3\AA & 3.75 & 78.0\% & \\
\midrule
C$\alpha$ & 5\AA & 3.20 & 80.9\% &  \\
C$\alpha$ \& C$\beta$ & 5\AA & 3.87 & 73.1\% & \\ 
All Atoms & 5\AA & 3.80 & 74.5\% &  \\
\midrule
C$\alpha$ & 8\AA & 3.80 & 74.5\% & \\
C$\alpha$ \& C$\beta$ & 8\AA & 3.87 & 73.8\% &  \\
All Atoms & 8\AA & 4.20 & 78.7\% &  \\
\bottomrule
\end{tabular}
\end{center}
\caption{\small Homogeneous reconstruction model accuracy for different settings of generating the dataset of cryo-EM images (resolution and which atoms are included). The model performs similarly across these choices. For other homogeneous experiments, we use all atoms at 3\AA. For heterogeneous experiments, we use $C_\alpha$ at 3\AA.}
\label{tab:homo_generative}
\end{table}

\subsection{cryoDRGN baseline}

For homogeneous reconstruction, we train a cryoDRGN positionally-encoded 3-layer MLP of width 256 for 20 epochs. For heterogeneous reconstruction, both the encoder and decoder networks are 3-layer MLPs of width 256, and are trained for 20 epochs.

\end{document}